\documentclass[%
 aip,
 amsmath,amssymb,
 reprint,%
]{revtex4-1}

\usepackage{graphicx}
\usepackage{dcolumn}
\usepackage{bm}

\usepackage[utf8]{inputenc}
\usepackage[T1]{fontenc}
\usepackage{mathptmx}
\usepackage{etoolbox}

\usepackage{xcolor}

\definecolor{deep_blue}{RGB}{31, 119, 180}
\definecolor{deep_orange}{RGB}{255, 127, 14}
\definecolor{deep_green}{RGB}{44, 160, 44}
\definecolor{deep_red}{RGB}{214, 39, 40}
\definecolor{deep_purple}{RGB}{148, 103, 189}
\definecolor{deep_brown}{RGB}{140, 86, 75}
\definecolor{deep_pink}{RGB}{227, 119, 194}

\usepackage[colorlinks,linktoc=all]{hyperref}
\hypersetup{
  linkcolor=deep_pink,    
  citecolor=deep_green,   
  urlcolor=deep_red       
}
\usepackage{cleveref}

\makeatletter
\def\@email#1#2{%
 \endgroup
 \patchcmd{\titleblock@produce}
  {\frontmatter@RRAPformat}
  {\frontmatter@RRAPformat{\produce@RRAP{*#1\href{mailto:#2}{#2}}}\frontmatter@RRAPformat}
  {}{}
}%
\makeatother
\begin{document}

\preprint{AIP/123-QED}

\title[Storage and selection of multiple chaotic attractors in minimal reservoir computers]{Storage and selection of multiple chaotic attractors in minimal reservoir computers}
  
\author{Francesco Martinuzzi}
    \affiliation{Max Planck Institute for the Physics of Complex Systems, Dresden, Germany}
    \email{martinuzzi@pks.mpg.de}

\author{Holger Kantz}
    \affiliation{Max Planck Institute for the Physics of Complex Systems, Dresden, Germany}

\date{\today}
\begin{abstract}
    Modern predictive modeling increasingly calls for a single learned dynamical substrate to operate across multiple regimes. From a dynamical-systems viewpoint, this capability decomposes into the storage of multiple attractors and the selection of the appropriate attractor in response to contextual cues.
    In reservoir computing (RC), multi-attractor learning has largely been pursued using large, randomly wired reservoirs, on the assumption that stochastic connectivity is required to generate sufficiently rich internal dynamics. At the same time, recent work shows that minimal deterministic reservoirs can match random designs for single-system chaotic forecasting.
    Under which conditions can minimal topologies learn multiple chaotic attractors?
    In this paper, we find that minimal architectures can successfully store multiple chaotic attractors. However, these same architectures struggle with task switching, in which the system must transition between attractors in response to external cues.
    We test storage and selection on all 28 unordered system pairs formed from eight three-dimensional chaotic systems. We do not observe a robust dependence of multi-attractor performance on reservoir topology. Over the ten topologies investigated, we find that no single one consistently outperforms the others for either storage or cue-dependent selection.
    Our results suggest that while minimal substrates possess the representational capacity to model coexisting attractors, they may lack the robust temporal memory required for cued transitions.
\end{abstract}

\maketitle

\begin{quotation}
    When modeling a physical phenomenon, simpler models are usually preferred because they are more tractable and interpretable. For machine learning models, simplicity entails fewer parameters and easier training. Reservoir computing (RC) models sit naturally at this intersection: the recurrent core is fixed a priori, so its structure (and thus its behavior) can be specified and analyzed directly, and its training is linear regression. Here, we ask how simple the reservoir structure can be while still learning multiple chaotic attractors. Supporting multiple behaviors splits broadly into two requirements: storage, meaning that multiple attractors can be represented within the same substrate, and selection, meaning that an external cue can reliably change the dynamics to the desired attractor. We evaluate the storage and selection of ten different minimal RC models for learning two distinct chaotic systems. We find that minimal deterministic reservoirs can often achieve attractor storage when trained to reproduce multiple systems in parallel, but they typically fail at cue-driven selection in a switching task. This separation suggests that representing multiple chaotic patterns does not necessarily require stochastic reservoirs, whereas reliable switching likely demands additional memory or control structure beyond the minimal designs studied here.
\end{quotation}

\section{\label{mmfrc:sec:intro} Introduction}

    Many scientific modeling problems require a single predictive model to operate across multiple dynamical regimes. For instance, weather conditions vary across geographic regions \citep{toth2026weather}, and robotic components may need to switch responses under changing external stimuli \citep{brooks1986robust}. In such settings, a model must store multiple distinct dynamical patterns and reliably deploy the appropriate one when conditions change. This ability can be seen as composed of two different mechanisms: \emph{storage}, meaning that multiple stable attractors can coexist in the same system, and \emph{selection}, meaning that context---or cues---reliably routes the dynamics to the desired attractor \citep{pisarchik2014control}. Biological brains naturally store different patterns by reusing fixed neural circuitry (neural reuse, or multifunctionality) \citep{anderson2014allocating, mccaffrey2015brain}. In machine learning (ML), analogous capabilities are increasingly pursued via foundation models (FMs) \citep{bommasani2021opportunity}. Built to internalize families of behaviours, FMs can be steered at inference time by prompts, conditioning variables, or other context signals. For example, time-series FMs can adapt their forecasts across regimes when provided with a context window representative of novel dynamics \citep{yeh2023toward, das2024decoder, ansari2024chronos}, and protein FMs can represent and generate diverse structural and functional motifs within a single parameterization \citep{nijkamp2023progen2}.

    Data-driven modeling of chaotic dynamics has also begun to shift toward architectures that can learn and switch between distinct attractors. Traditionally, models were trained on a single chaotic system within a fixed-parameter regime to either recover short-term trajectories or reproduce long-term statistical properties \citep{cao1995predicting, bakker2000learning, vlachas2018data, chattopadhyay2020data, li2022learning, hess2023generalized}. However, recent advances demonstrate that ML systems can generalize across different dynamical regimes. Modern approaches can, for instance, recover trajectories in unseen portions of the phase space \citep{goring2024outofdomain, norton2025learning}, switch between different dynamical regimes in the same system \citep{kim2021teaching}, and anticipate critical transitions and tipping points \citep{kong2021machine, xiao2021predicting, huang2024deeplearning}. Furthermore, general-purpose time-series FMs already exhibit an emergent ability to forecast chaotic systems with no additional training \citep{zhang2025zeroshot}. Building on these results, new domain-specific FMs are now being developed specifically to generalize multiple chaotic systems, enabling a single architecture to generate diverse, mathematically faithful attractors at prediction time \citep{hemmer2025true, lai2025panda}.

    Within this line of research, reservoir computing (RC) has emerged as a particularly flexible framework. In RC, the recurrent “reservoir” dynamics are randomly initialized and fixed, while only the readout is trained with linear regression \citep{verstraeten2005reservoir, verstraeten2007experimental}. The internal reservoir can hence be treated as an explicit dynamical substrate whose dimensionality, connectivity, and weight structure are directly tunable. The combination of simple, fast training and additional control over reservoir properties makes RC models a valuable framework for mechanistic exploration \citep{yan2024emerging}: architectural and dynamical ingredients can be varied while keeping the evolution equations consistent. Models such as echo state networks (ESNs) \citep{jaeger2001echo} have been used to model chaotic dynamics with great success \citep{jaeger2004harnessing, pathak2018modelfree, smith2022learning}. Extensions of ESNs have also shown the ability to learn multiple distinct chaotic systems; by changing the training algorithm, \citet{lu2020invertible} show how ESNs can switch between systems, thereby showcasing different dynamical regimes. Further studies demonstrate how ESNs can retrieve chaotic systems previously stored in memory \citep{kong2024reservoir}, and learn multistable systems \citep{roy2022modelfree}. Multiple dynamics can also be stored in RC systems using conceptors \citep{jeager2017using}. Recent works illustrate the ability of  ESNs to store multiple attractors in memory \citep{flynn2021multifunctionality, du2025multifunctional}. Follow-up investigations have used this multistable structure to probe bifurcations and transitions between learned attractors in reservoir models \citep{flynn2023seeing, flynn2024exploring, ohagan2025confabulation}, and similar multifunctional dynamics have been demonstrated in other RC architectures as well \citep{flynn2022exploring, morra2023multifunctionality, terajima2025multifunctional}.
        
    However, most RC approaches for multi-attractor learning still rely on large, stochastically constructed reservoirs. The structure of the reservoir adjacency matrix (also called its \emph{topology}) is known to strongly influence performance \citep{emmertstreib2006influence, carroll2019network, dale2019role, jaurigue2024chaotic, rathor2025asymmetrically, ylmazbingl2025reservoir, tangerami2025optimizing, rathor2025prediction}. Random sparse reservoirs, the default choice in ESNs, are explicitly motivated by the goal of producing a rich set of internal dynamics \citep{lukoeviius2009reservoir}, which are widely regarded as essential for reservoir computation both in single \citep{enel2016reservoir} and multiple \citep{kong2024reservoir} system settings. Recent evidence, however, suggests that this level of complexity is not a prerequisite for learning chaotic dynamics. Small reservoirs \citep{jaurigue2024chaotic}, sparsely connected architectures \citep{griffith2019forecasting}, and deterministic designs \citep{ma2023novel, ma2023efficient, viehweg2025deterministic} often provide performance comparable to large random reservoirs. In particular, minimal deterministic topologies offer more accurate and consistent results when modeling chaotic systems \citep{martinuzzi2025minimal}. Is it also possible for minimal topologies to store and recall multiple chaotic attractors? 

    In this paper, we evaluate whether minimal deterministic reservoir topologies can learn multiple chaotic systems in a single model. We consider two distinct approaches: the first is based on the blending technique (BT), in which a single ESN is trained to learn and replicate multiple systems \emph{at the same time} \citep{flynn2021multifunctionality}; the second is the parameter-aware (PA) approach, in which a single ESN can learn to \emph{switch} between different learned systems in response to an external cue signal \citep{panahi2024adaptable}. For each approach, we train a single ESN topology to learn the dynamics of pairs of chaotic systems drawn from a pool of eight benchmark systems, yielding 28 distinct experiments per setting. Exploring a broad set of system pairs reduces the risk that our conclusions depend on an atypical or “unlucky” combination of systems. We study ten simple deterministic reservoir topologies that collectively cover a broad range of designs proposed in previous work on deterministic reservoirs \citep{rodan2011minimum, rodan2012simple, elsarraj2019demystifying, fu2023doublecycle}. A topology is considered successful if a trained ESN can accurately reproduce both chaotic attractors associated with a given system pair. Under the BT protocol, success means that the ESN simultaneously generates faithful forecasts for both systems from the state input. Under the PA protocol, success requires cue-dependent selection of different dynamics. In our experiments, minimal deterministic topologies consistently achieve attractor storage in the BT setting, but they do not support robust cue-dependent selection in the PA setting.
    
    The structure of the paper is as follows. In Sec \ref{mmfrc:sec:meth} we detail the technical aspects of the work, starting from a description of ESNs and their minimal topologies in Sec. \ref{mmfrc:subsec:esn}, and how to train them to learn multiple systems in Sec. \ref{mmfrc:subsec:mf}. Further sections detail the metric used to quantify the validity of the results in Sec. \ref{mmfrc:subsec:mea} and the systems and preprocessing used in the study in Sec. \ref{mmfrc:subsec:data}. We present the results in Sec. \ref{mmfrc:sec:res}. Section \ref{mmfrc:sec:disc} provides a final summary and discussion of the work.


\begin{figure*}[ht]
    \centering
    \includegraphics[width=\textwidth]{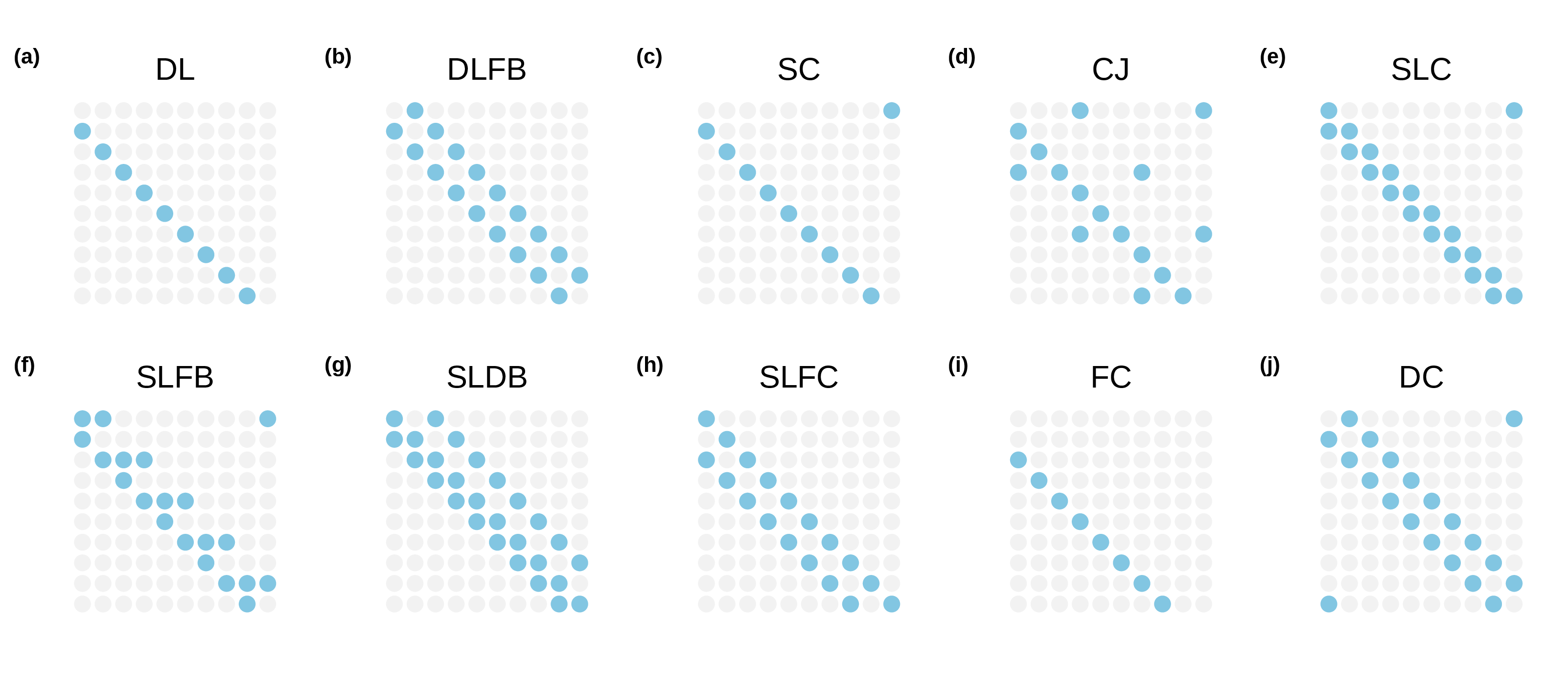}
    \caption{\textbf{Minimal deterministic reservoir topologies.} Each panel shows the nonzero entries (magenta) of the $10\times10$ reservoir matrix $\mathbf{W}$. All the nonzero weights have the same magnitude and sign. Zero entries are shown in light gray. Panels correspond to the following structures: (a) delay line (DL), (b) delay line with feedback connections (DLFB), (c) simple cycle (SC), (d) cycle with jumps (CJ), (e) self-loop cycle (SLC), (f) self-loop feedback cycle (SLFB), (g) self-loop delay line with backward connections (SLDB), (h) self-loop with forward connections (SLFC), (i) forward connections (FC), and (j) double cycle (DC).}
    \label{mmfrc:fig:minits}
\end{figure*}

\section{\label{mmfrc:sec:meth} Methods}

    \subsection{\label{mmfrc:subsec:esn} Minimal Reservoir Computers}

        RC refers to a family of ML models that uses dynamical systems for computation \citep{verstraeten2005reservoir, verstraeten2007experimental}. The shared component of RC architectures is the reservoir, a high-dimensional dynamical system. The role of the reservoir is to provide a nonlinear expansion of the data, exposing linear dependencies. Modeling with an RC involves three main stages: (i) the data is first passed through the RC model. (ii) Afterwards, the resulting reservoir \emph{states} are used for training against the desired target data. Linear regression is usually used for training in RC models \citep{jaeger2002tutorial}. Finally (iii), the model can be used for the chosen task. Any ML task, such as regression, forecasting, or classification, can be done with a reservoir computer.

        Amongst RC models, echo state networks (ESNs)\citep{jaeger2001echo} have long been studied in the field of nonlinear dynamics due to their simple construction and fast training. Stemming from the ML literature, ESNs use a recurrent neural network (RNN) as the reservoir. Unlike traditional RNNs, which are trained via backpropagation, ESNs' internal weights are randomly initialized and held fixed. Only the last layer is trained, using linear regression. Formally, let $\mathbf{u}(t) \in \mathbb{R}^{\text{D}_{\text{in}}}$ be the input vector at time t, then the time evolution of the ESN is represented by
        \begin{equation}
        \label{mmfrc:eq:esn}
            \mathbf{x}(t) = (1 - \alpha) \mathbf{x}(t) + \alpha \tanh\left(\mathbf{W}_\text{in} \mathbf{u}(t) + \mathbf{W} \mathbf{x}(t-1) + \mathbf{b} \right),
        \end{equation}
        where $\mathbf{x}(t) \in \mathbb{R}^{\text{D}_{\text{res}}}$ is the reservoir state at time $t$, and $\alpha$ is the leaky coefficient, which controls the timescales of the evolution of the ESN. The matrix $\mathbf{W}_\text{in} \in \mathbb{R}^{\text{D}_{\text{res}} \times \text{D}_{\text{in}}}$ denotes the \emph{input} matrix, while $\mathbf{W} \in \mathbb{R}^{\text{D}_{\text{res}} \times \text{D}_{\text{res}}}$ is usually referred to as the \emph{reservoir} matrix. The term $\mathbf{b} \in \mathbb{R}^{\text{D}_{\text{res}}}$ adds a bias component, useful to break internal simmetries \citep{}.

        Because of the random and static nature of ESN matrices, their initialization plays an important role. In traditional ESN literature, the input and reservoir matrices are initialized randomly, with various hyperparameters controlling their overall behavior. General guidelines dictate that the input matrix be randomly generated with weights drawn from $\mathcal{U}(-\sigma, \sigma)$, where $\sigma$ controls the nonlinearity that is applied to the input, and is subsequently treated as a hyperparameter. The reservoir matrix requires greater care, as it is the primary driver of the ESN's dynamics. This matrix is generally generated as sparse with sparsity $\theta$, following a Erd\H{o}s--R\'enyi connection graph. After being generated, the matrix $\mathbf{W}^*$ is then scaled to a chosen spectral radius $\rho$ to obtain the final reservoir matrix $\mathbf{W}$. It follows that $\rho$ and $\theta$ are considered hyperparameters to be tuned. More specifically, the spectral radius plays an important, and still hotly debated, role in the performance of the ESN \citep{jiang2019modelfree, hart2023estimating, hart2024attractor}.
        
        To reduce this complexity and assess whether it is necessary in the first place, recent approaches have used simpler setups for building ESNs. First proposed by \citet{rodan2011minimum}, minimum-complexity ESNs (MESNs) offer a simple approach to ESN construction. The rules follow simple steps: (i) all the weights used are the same, with their magnitude set and not randomly defined. (ii) The topologies of the reservoirs are also set and not random. (iii) the signs of the reservoir weights are the same, while the signs of the weights of the input layer do change. Following these rules, multiple alternative topologies have been proposed \citep{rodan2012simple, elsarraj2019demystifying, fu2023doublecycle}. In this work, we consider the following topologies:
        \begin{itemize}
            \item \textbf{Delay line (DL)}\citep{rodan2011minimum}, composed only of weights arranged in a line. The nonzero entries of the matrix are located on the lower subdiagonal $\mathbf{W}^{(i+1,i)} = r, i \in [1, \text{D}_{res}-1]$.
            
            \item \textbf{Delay line with feedback connections (DLFB)}\citep{rodan2011minimum}, which shares the same structure as DL, with the addition of feedback weights. The matrix is built with nonzero weights on the lower subdiagonal $\mathbf{W}^{(i+1,i)} = r$ and on the upper subdiagonal $\mathbf{W}^{(i,i+1)} = b$ where $i \in [1, \text{D}_{res}-1]$.
                        
            \item \textbf{Simple cycle (SC)}\citep{rodan2011minimum}, where the weights form a cycle. The SC matrix has nonzero elements on the lower subdiagonal $\mathbf{W}^{(i+1,i)} = r, i \in [1, \text{D}_{res}-1]$, with the addition of a weight located in the upper right corner $\mathbf{W}^{(1,\text{D}_{res})} = r$

            \item \textbf{Cycle with jumps (CJ)}\citep{rodan2012simple}, builds on SC, but adds bidirectional jump connections of fixed distance $\ell$, all with weight $r_j$. Nonzero elements of CJ are on the lower subdiagonal $\mathbf{W}^{(i+1,i)} = r, \; i \in [1, \text{D}_{\text{res}}-1]$, and the upper right corner $\mathbf{W}^{(1,\text{D}_{\text{res}})} = r$, following SC. Additionally, jump entries $r_j$ are added, with a jump size $1 < \ell < \lfloor \text{D}_{\text{res}}/2 \rfloor$. If $(\text{D}_{\text{res}} \bmod \ell) = 0$, there are $\text{D}_{\text{res}} / \ell$ jumps. The first jump starts from the first weight in the cycle, and connects it to $1+\ell$. The last jump is from unit $\text{D}_{\text{res}}+1-\ell$ to the first one. If $(\text{D}_{\text{res}} \bmod \ell) \neq 0$, there are $\lfloor \text{D}_{\text{res}} / \ell \rfloor$ jumps, the last ending in unit $\text{D}_{\text{res}}+1-(\text{D}_{\text{res}} \bmod \ell)$. The jumps are bidirectional and share the same connection weight $r_j$.

            \item \textbf{Self-loop cycle (SLC)}\citep{elsarraj2019demystifying}, also builds on the SC with the addition of self-loops. Nonzero weights are $\mathbf{W}^{(i+1,i)} = r, i \in [1, \text{D}_{res}-1]$, $\mathbf{W}^{(1,\text{D}_{res})} = r$, and $\mathbf{W}^{(i,i)} = ll, i \in [1, \text{D}_{res}]$

            \item \textbf{Self-loop feedback cycle (SLFB)}\citep{elsarraj2019demystifying}, continues to build on the cycle reservoir, with the addition of self-loops and feedbacks. Similarly to the SC, nonzero weights are $\mathbf{W}^{(i+1,i)} = r, i \in [1, \text{D}_{res}-1]$, $\mathbf{W}^{(1,\text{D}_{res})} = r$. Additionally, if $i$ is odd $\mathbf{W}^{(i,i)} = ll, i \in [1, \text{D}_{res}]$, otherwise $\mathbf{W}^{(i,i+1)} = r, i \in [1, \text{D}_{res}-1]$.

            \item \textbf{Self-loop delay line with backward connections (SLDB)}\citep{elsarraj2019demystifying}, extends the delay-line by adding self-loops on every unit and backward links to the second previous unit. Nonzero entries are $\mathbf{W}^{(i,i)} = ll$ for $i \in [1,\text{D}_{\text{res}}]$ (self-loops), $\mathbf{W}^{(i+1,i)} = r$ for $i \in [1,\text{D}_{\text{res}}-1]$ (forward path), and $\mathbf{W}^{(i,i+2)} = r$ for $i \in [1,\text{D}_{\text{res}}-2]$ (backward connections to the second previous unit).
            
            \item \textbf{Self-loop with forward connections (SLFC)}\citep{elsarraj2019demystifying}, removes the standard forward path and keeps only connections of length two, forming two disjoint forward chains over odd and even indices, plus self-loops on every unit. The nonzero entries are $\mathbf{W}^{(i,i)} = ll$ for $i \in [1,\text{D}_{\text{res}}]$ and $\mathbf{W}^{(i+2,i)} = r$ for $i \in [1,\text{D}_{\text{res}}-2]$.
            
            \item \textbf{Forward connections (FC)}\citep{elsarraj2019demystifying}, identical to SLFC but without self-loops. The only nonzero entries are $\mathbf{W}^{(i+2,i)} = r$ for $i \in [1, \text{D}_{\text{res}}-2]$.

            \item \textbf{Double cycle (DC)}\citep{fu2023doublecycle}, again builds on SC, but it adds an additional cycle in the opposite direction. The nonzero elements are then $\mathbf{W}^{(i+1,i)} = r, i \in [1, \text{D}_{res}-1]$ and $\mathbf{W}^{(1,\text{D}_{res})} = r$ as in SC, plus other cycle $\mathbf{W}^{(i,i+1)} = r, i \in [1, \text{D}_{res}-1]$ and $\mathbf{W}^{(\text{D}_{res}, 1)} = r$
        \end{itemize}
        To further minimize the number of free parameters in these initializers, we set $r=ll=r_j=0.1$ in this work.

        \paragraph{\label{mmfrc:subsubsec:tr} Training} Given an input sequence $\{\mathbf{u}(t)\}_{0}^{T}$, we drive the reservoir using Eq.~\ref{mmfrc:eq:esn}. Following standard procedures, we remove an initial transient period $\tau = 300$ from the training states, resulting in a set of states $\{\mathbf{x}(t)\}_{\tau}^{T}$. Each state $\mathbf{x}(t)$ is subsequently transformed with a function such that $\mathbf{h}(t) = \mathbf{\mathit{H}}(\mathbf{x(t)}) = [\mathbf{x}(t); \mathbf{x}^2(t)]$, where $[;]$ indicates vertical concatenation. The resulting states are collected into a states matrix $\mathbf{H}$, where each column corresponds to one time step. We likewise collect the desired outputs into a target matrix $\mathbf{Y}_{\mathrm{target}}$. Training then consists of fitting a linear readout $\mathbf{W}_{\mathrm{out}}$ that maps reservoir states to targets. We estimate $\mathbf{W}_{\mathrm{out}}$ via ridge regression by minimizing
        \begin{equation}
            \label{mmfrc:eq:rr}
            \mathcal{L}(\mathbf{W}_{\text{out}}) 
            = \left\| \mathbf{Y}_{\text{target}} 
              - \mathbf{W}_{\text{out}} \mathbf{H} \right\|_2^2 
              + \lambda \left\| \mathbf{W}_{\text{out}} \right\|_2^2.
        \end{equation}
        The minimizer of Eq.~\ref{mmfrc:eq:rr} admits the closed-form solution
        \begin{equation}
            \label{mmfrc:eq:rrcf}
            \mathbf{W}_{\text{out}}
            = \mathbf{Y}_{\text{target}} \mathbf{H}^\top
              \left( \mathbf{H}\mathbf{H}^\top + \lambda \mathbf{I} \right)^{-1},
        \end{equation}
        where $\mathbf{I}$ is the identity matrix, and $\lambda$ is the regularization coefficient controlling the strength of the $\ell_2$ penalty.

        \paragraph{\label{mmfrc:subsubsec:fr} Forecasting} After estimating $\mathbf{W}_{\mathrm{out}}$, we run the ESN in forecasting mode by computing the output
        \begin{equation}
            \label{mmfrc:eq:fc}
            \mathbf{v}(t) = \mathbf{W}_{\text{out}} \mathbf{h}(t),
        \end{equation}
        where $\mathbf{v}(t)$ denotes the ESN output at time $t$. Throughout this work, we evaluate models in an autoregressive (closed-loop) forecasting regime. During training, the target sequence is chosen one step ahead of the input, i.e., $\mathbf{y}_{\mathrm{target}}(t)=\mathbf{u}(t+1)$, so that the trained readout approximates the one-step map $\hat{\mathbf{u}}(t+1)=\mathbf{v}(t)$. At prediction time, the ESN is run autonomously by feeding this estimate back as the next input and iterating for $T$ steps. Under this closed-loop rollout, Eq.~\eqref{mmfrc:eq:esn} becomes
        \begin{equation}
        \label{mmfrc:eq:esnpred}
            \mathbf{x}(t) = (1 - \alpha) \mathbf{x}(t) + \alpha \tanh\left(\mathbf{W}_\text{in} \mathbf{W}_\text{out} \mathbf{x}(t-1) + \mathbf{W} \mathbf{x}(t-1) + \mathbf{b} \right),
        \end{equation}
        \begin{figure*}[ht]
            \centering
            \includegraphics[width=\textwidth]{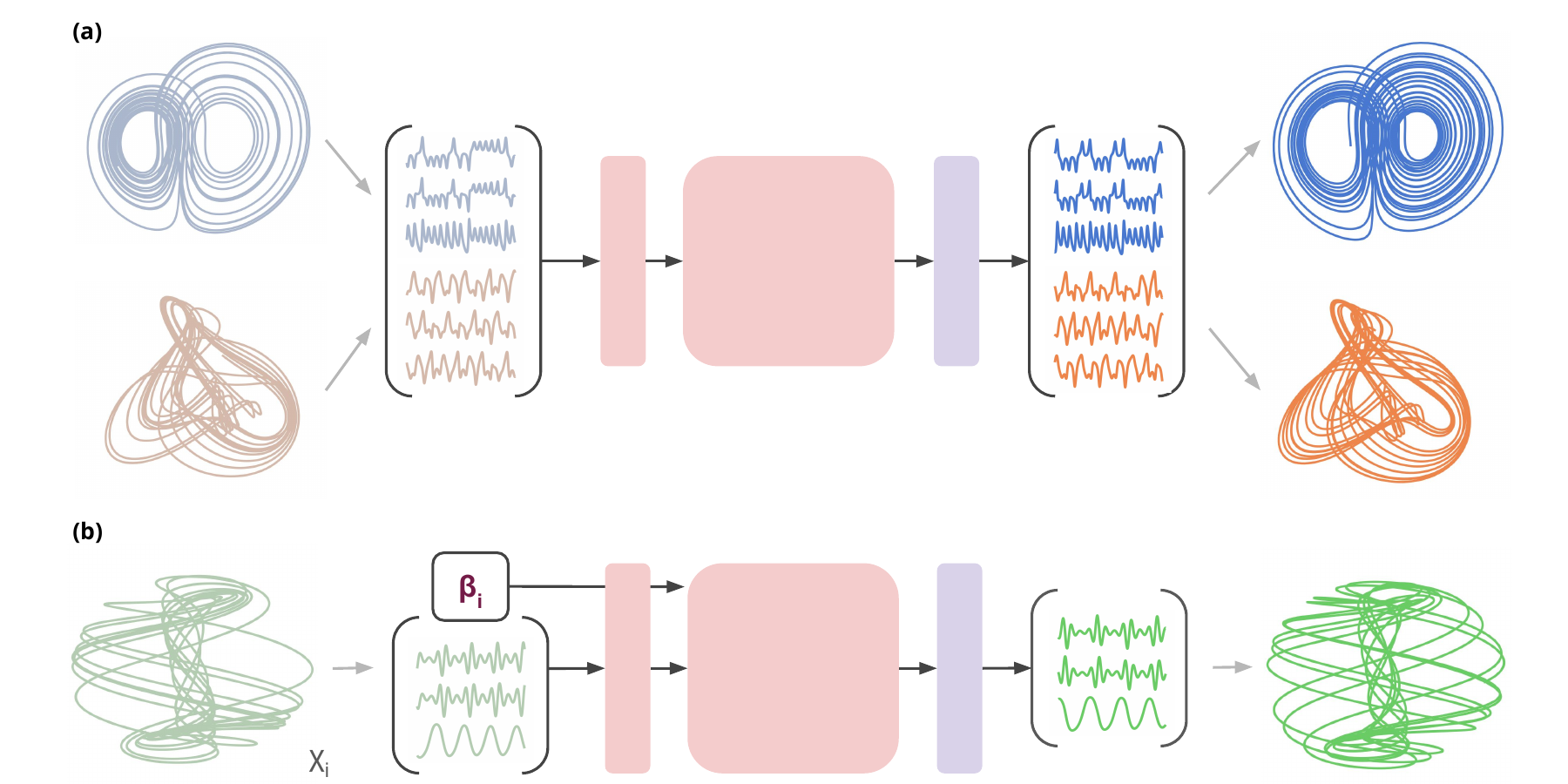}
            \caption{\textbf{Approaches for multi-attractor learning in reservoir computing.} Panel (a) shows the approach for attractor storage in an echo state network (ESN) using the blending technique (BT). Two chaotic trajectories are provided in parallel by concatenating their state vectors into a single input, \mbox{$\mathbf{u}_{\mathrm{bt}}(t)=[\mathbf{u}_{\chi_1}(t);\mathbf{u}_{\chi_2}(t)]$}. This joint input drives the ESN, and a single linear readout is trained to produce a concatenated output trajectory that reconstructs both systems simultaneously (right time series and attractors; colors distinguish the two systems). Panel (b) illustrates cue-dependent selection using the parameter-aware (PA) approach. In addition to the state input, a scalar cue $\beta_i$ identifying the target attractor is injected through the bias term in Eq.~\eqref{mmfrc:eq:paesn}. The ESN is then run with $\beta_i$ held fixed so that the autonomous dynamics are expected to converge to, and reproduce, the attractor associated with that cue. In both panels, the light pink boxes indicate untrained matrices, while the light violet box indicates the trained readout.}

            \label{mmfrc:fig:mf}
        \end{figure*}

    \subsection{\label{mmfrc:subsec:mf} Protocols for multi-attractor learning}

        Since standard ESNs are trained for a single input--output mapping, the setup of Sec.~\ref{mmfrc:subsec:esn} does not, by itself, address the multi-attractor setting studied here. While there exist different ways to learn multiple dynamics in a reservoir computing context (see the relevant paragraph in Sec~\ref{mmfrc:sec:intro}), in this work we limit our exploration to two distinct methods. Specifically, we consider (a) the blending technique (BT; Sec.~\ref{mmfrc:subsubsec:bt}), which probes attractor \emph{storage} by training on concatenated trajectories, and (b) the parameter-aware (PA; Sec.~\ref{mmfrc:subsubsec:pa}) protocol, which tests \emph{cue-dependent selection} by injecting a label signal that is intended to route the dynamics toward a chosen attractor. Before passing the attractors to the ESNs, we separate them in phase space. Each attractors' states are shifted by a value $\eta = 0.2$. In our case of two distinct attractors, the coordinates are shifted as follows: $(x + \eta, y + \eta, z + \eta)$, $(x - \eta, y - \eta, z - \eta)$.

        \subsubsection{\label{mmfrc:subsubsec:bt} Blending Technique}

            Proposed by \citet{flynn2021multifunctionality}, in the BT approach, the ESN is fed with a concatenation of the evolution vectors of two attractors $\chi_i$, $i \in \{1,2\}$. Let $\mathbf{u}_{\chi_1}(t) \in \mathbb{R}^{D_{\text{in},1}}$ be the coordinates of the points of attractor $\chi_1$, and $\mathbf{u}_{\chi_2}(t) \in \mathbb{R}^{D_{\text{in},2}}$ be the coordinates of the points of attractor $\chi_2$. The ESN input vector is then
            \begin{equation}
            \label{mmfrc:eq:bt}
                \mathbf{u}_{\text{bt}}(t) =
                \begin{bmatrix}
                    \mathbf{u}_{\chi_1}(t) \\
                    \mathbf{u}_{\chi_2}(t)
                \end{bmatrix},
            \end{equation}
            where $\mathbf{u}_{\text{bt}}(t) \in \mathbb{R}^{D_{\text{in},1} + D_{\text{in},2}}$. In the context of this work we use $D_{\text{in},1} = D_{\text{in},2} = 3$. Figure~\ref{mmfrc:fig:mf}a shows a schematic representation of how an RC is trained in the BT technique.
            
            We note that, in the original paper, an additional parameter $\beta \in [0,1]$ controls the \emph{blending} (hence the naming),
            \begin{equation}
                \mathbf{u}_{\text{bt}}(t) =
                \begin{bmatrix}
                    \beta \, \mathbf{u}_{\chi_1}(t) \\
                    (1-\beta) \, \mathbf{u}_{\chi_2}(t)
                \end{bmatrix},
            \end{equation}
            providing a bifurcation parameter for the multifunctional ESN. In this work, we use concatenation without the blending parameter to focus strictly on determining whether minimal ESNs can store multiple attractors. Removing the blending parameter does not impact the performance of the models, as it is implied by Fig. 7a of Ref.~\onlinecite{flynn2021multifunctionality}.

        \subsubsection{\label{mmfrc:subsubsec:pa} Parameter Aware Technique}

            Similarly to \citet{ohagan2025confabulation}, we implement a simplified version of the PA technique for training ESNs \citep{kong2021machine, xiao2021predicting, kong2024reservoir}. In this technique, each attractor is identified by assigning it a unique scalar label. The attractor label is then fed into the ESN via a parameter-dependent bias term in the state update to recover the associated dynamics.
                       
            Formally, let $\beta_i \in \mathbb{R}$, $i \in \{1,\dots,m\}$, be the scalar label for each of the $m$ systems. The parameter-aware ESN is then defined as
            \begin{equation}
            \label{mmfrc:eq:paesn}
                \mathbf{x}(t) = (1 - \alpha) \mathbf{x}(t-1) + \alpha \tanh\left(\mathbf{W}_\text{in} \mathbf{u}_{\beta_{i}}(t) + \mathbf{W} \mathbf{x}(t-1) + \beta_{i}\mathbf{b} \right).
            \end{equation}
            where $\mathbf{u}(t)$ is the input at time $t$ and $\beta_i$ is kept fixed for all samples belonging to attractor $i$. We depict the labeling scheme of PA in Fig.~\ref{mmfrc:fig:mf}. While the figure shows the RC being trained with a single system, the method supports storing multiple attractors, each with its own label.
            
            The difference in the values of $\beta_i$ between attractors is governed by the hyperparameter $\delta\beta$, which controls the spacing between labels via $\beta_i = \beta_{i-1} + \delta\beta$. Following preliminary hyperparameter selection, we choose $\delta\beta = 0.4$, starting from $\beta_1 = 0.1$ for the first attractor.
            
            Training follows the standard ESN procedure, with the addition that the label $\beta_i$ is provided alongside each training sample and enters the state update through the bias term in Eq.~\eqref{mmfrc:eq:paesn}. Concretely, for each system $i\in\{1,\dots,m\}$ we generate a training trajectory $\{\mathbf{u}_i(t)\}_{t=1}^{T}$ and associate every sample in that trajectory with the constant label $\beta_i$. We then construct a single training sequence by concatenating these labeled trajectories in time. During this concatenated teacher-forced pass, the input $\mathbf{u}(t)$ is the current true state from the active system segment and the label is set to the corresponding value $\beta(t)=\beta_i$ for that entire segment; when the sequence transitions to the next system segment, the label is switched accordingly. Reservoir states are collected from this full labeled sequence and a single readout $\mathbf{W}_{\mathrm{out}}$ is trained by ridge regression as in the standard ESN.

            During autoregressive forecasting, we select the target system by fixing the label to $\beta(t)\equiv\beta_i$ for all prediction steps and initializing the ESN with suitable initial conditions. The network is then run in closed loop: at each step the predicted output $\hat{\mathbf{u}}(t+1)$ is fed back as the next input, yielding the autonomous update in Eq.~\eqref{mmfrc:eq:esnpred} with the bias term scaled by the chosen label $\beta_i$. The accuracy of the forecast is assessed by evaluating all reproduced attractors individually, and then reporting the average score.

            \begin{figure*}[ht]
            \centering
            \includegraphics[width=\textwidth]{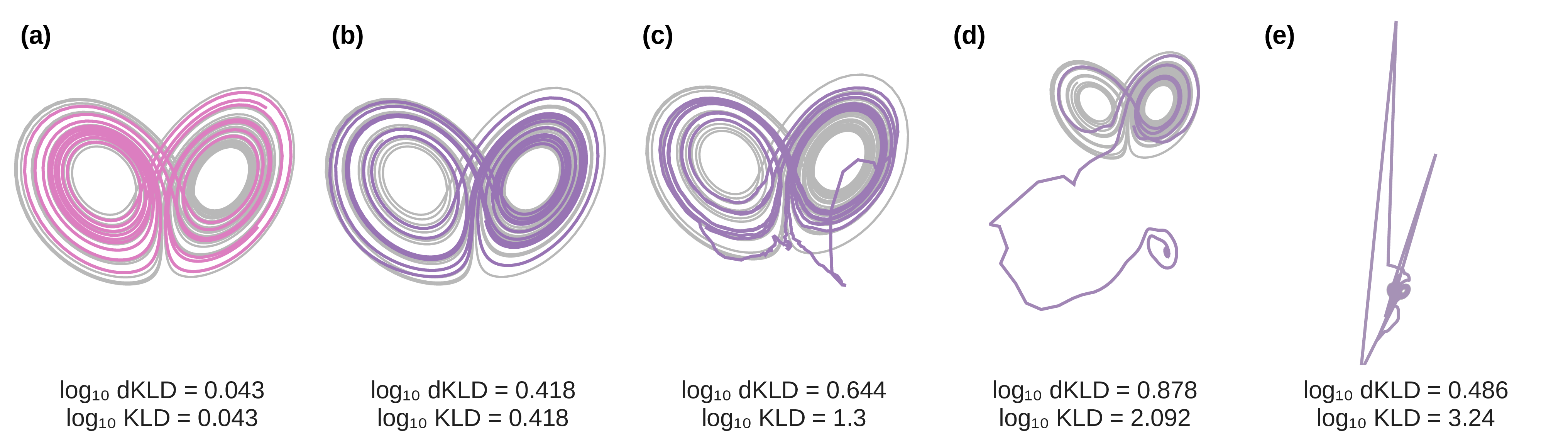}
            \caption{\textbf{Effect of prediction deterioration on metric changes.} In this figure, we provide Lorenz attractor overlays, illustrating how reconstruction quality maps to the discrete state--space Kullback--Leibler divergence (KLD). Grey curves show the reference Lorenz trajectory. In panel (a), the comparison trajectory (pink) is generated by the same Lorenz equations but using different initial conditions; both the standard discrete KLD (dKLD) and the penalized variant (KLD) remain small. Panels (b--e) show progressively degraded ESN forecasts (violet). As prediction quality deteriorates and forecasted trajectories increasingly deviate from the true state space, the standard dKLD exhibits relatively low sensitivity. On the other hand, the penalized KLD increases with the degree of prediction failure, providing a clearer indication of reconstruction failure. Metric values reported below each panel are in $\log_{10}$ scale.}
            \label{mmfrc:fig:metric}
        \end{figure*}

    \subsection{\label{mmfrc:subsec:mea} Accuracy metric - Kullback-Leibler divergence}

        As a measure of model performance, we use the Kullback-Leibler divergence (KLD) defined across space, in line with previous works \citep{wood2010statistical, koppe2019identifying, brenner2022tractable, platt2023constraining}. The chosen metric ensures that the predicted attractor aligns geometrically with the true attractor. Following \citet{hemmer2025true} we discretize the $\text{D}_{\text{in}}=3$-dimensional state space into $K = m^{N}$ bins using $m=30$ bins per dimension. Let $\hat{p}^{\mathrm{true}}_{i}$ and $\hat{p}^{\mathrm{RC}}_{i}$ denote the normalized occupancy frequencies of bin $i$ estimated from the ground-truth trajectory and from the reservoir-generated trajectory, respectively. The dKLD of a given RC per model is then approximated as 
        \begin{equation}
            \mathrm{dKLD}
            \;\approx\;
            \sum_{i=1}^{K} \hat{p}^{\mathrm{true}}_{i}\,
            \log\!\left(\frac{\hat{p}^{\mathrm{true}}_{i}}{\hat{p}^{\mathrm{RC}}_{i}}\right).
            \label{eq:kld_discrete}
        \end{equation}
        However, the metric in Eq.~\ref{eq:kld_discrete} only compares occupancy probabilities on the finite grid defined over the region spanned by the \emph{true} attractor. In practice, reservoir predictions may drift outside this reference domain. The corresponding excursions would then not be reflected by the histogram-based dKLD. To account for catastrophic failures, we augment the divergence with an out-of-bounds (OOB) penalty. Concretely, we define an axis-aligned bounding box from the true trajectory by setting, for each state-space dimension $d$, lower and upper limits $\ell_d = \min(x^{\mathrm{true}}_d) - 0.1\,\sigma^{\mathrm{true}}_d$ and $h_d = \max(x^{\mathrm{true}}_d) + 0.1\,\sigma^{\mathrm{true}}_d$, where $\sigma^{\mathrm{true}}_d$ is the standard deviation of the true coordinates along dimension $d$. We then compute the maximum relative violation of the predicted trajectory with respect to this box,
        \begin{equation}
            v \;=\; \max_{d}\,\frac{\max\!\bigl(0,\,\ell_d-\min(x^{\mathrm{pred}}_d)\bigr)+\max\!\bigl(0,\,\max(x^{\mathrm{pred}}_d)-h_d\bigr)}{h_d-\ell_d},
        \end{equation}
        and, whenever $v>0$, we return a penalty value
        \begin{equation}
            \mathrm{KLD} \;=\; w_{\mathrm{OOB}}\, v,
        \end{equation}
        with weight $w_{\mathrm{OOB}}=10^{2}$. Otherwise ($v=0$), we proceed with the histogram-based dKLD computation on the binning induced by $[\ell_d,h_d]$. The modified KLD ensures that predictions that leave the support of the true attractor are explicitly penalized. The resulting KLD values will also appear much larger than those reported in similar works. Figure~\ref{mmfrc:fig:metric} illustrates the difference between the two metrics for varying levels of attractor reconstruction quality. It is possible to observe that the KLD with OOD penalty returns results that are consistent with the visible deterioration in the attractor reconstruction by the ESN.

        \begin{figure*}[ht]
            \centering
            \includegraphics[width=\textwidth]{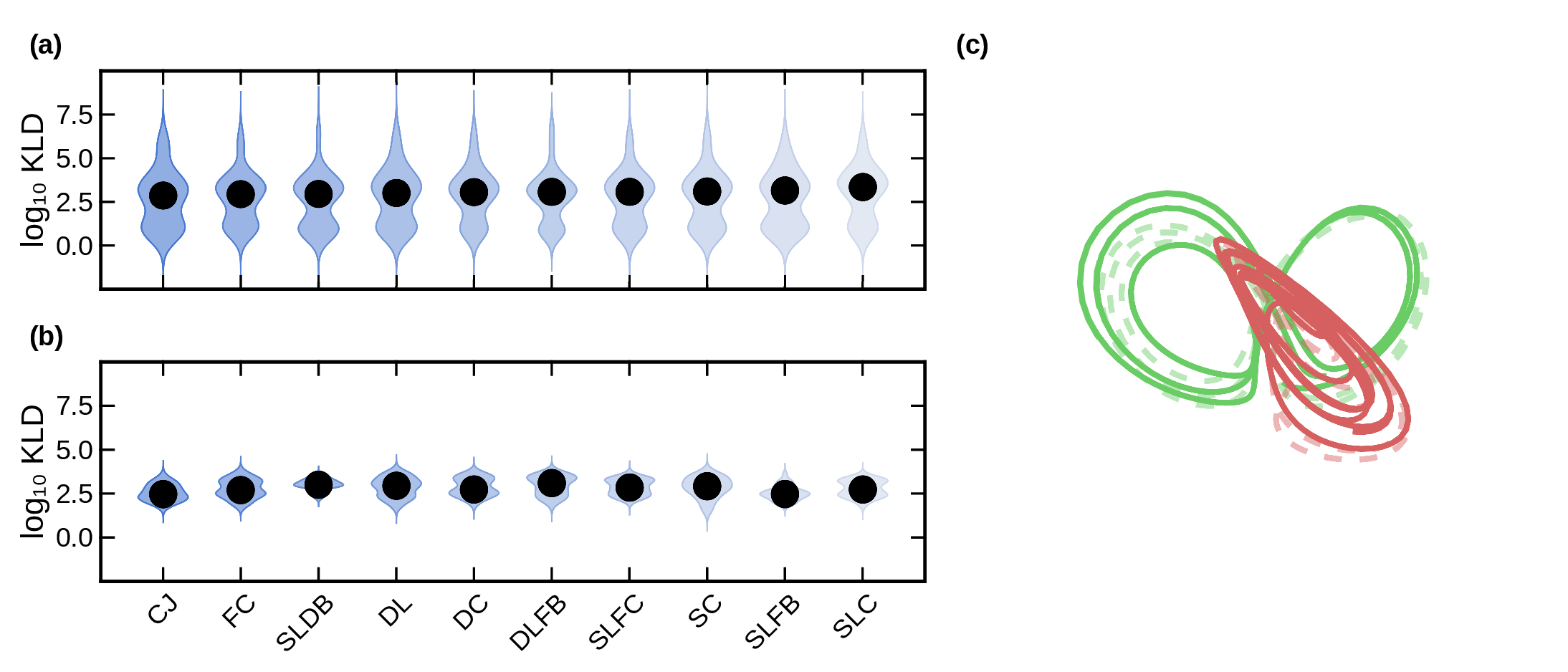}
            \caption{\textbf{Multi-attractor learning accuracy across minimal reservoir topologies}. Panel (a) shows the results of the blending technique (BT). The violin plots indicate the distributions of $\log_{10}$ Kullback-Leibler divergence (KLD) over all 28 unordered system pairs for each reservoir topology (ordered by increasing median). Panel (b) illustrates the results for the parameter-aware (PA) approach. The panel maintains the same topology ordering and visualization used in BT. In both panels, violins show the distribution across system pairs, and the black dot marks the median $\log_{10}\mathrm{KLD}$ for that topology; the apparent extension below zero is due to violin smoothing and does not imply negative KLD values. (c) Representative system pair reconstruction example for Lorenz–SprottS using the self-loop delay line with backward connections (SLDB) topology in the BT setting. Dashed curves show the reference attractors and solid curves the ESN-generated attractors, with different colors distinguishing the two systems. The accuracy for the system pair corresponds to $\log_{10}\mathrm{KLD}=0.692$.}
            \label{mmfrc:fig:mainkld}
        \end{figure*}

    \subsection{\label{mmfrc:subsec:data} Chaotic systems and data generation}

        We consider eight different chaotic systems taken from \texttt{dysts} \citep{gilpin2021chaos}. From this set, we form all unique combinations of two distinct systems. This yields $\binom{8}{2} = 28$ unique two-system combinations, which we refer to as system pairs. Each combination defines the systems that the minimal reservoir computers are tasked to reproduce. We provide the list of the systems used (with respective equations, parameters, and integration steps) in Appendix \ref{mmfrc:app:cs}. Each system trajectory is normalized to have a mean of zero and a standard deviation of 1 \citep{lu2017reservoir}.
        
        For the numerical integration, we sourced the system definitions from the Julia port of \texttt{dysts},  \texttt{ChaoticDynamicalSystemLibrary.jl}\footnote{\url{https://github.com/nathanaelbosch/ChaoticDynamicalSystemLibrary.jl}}. We use the \texttt{Feagin12} solver from the \texttt{DifferentialEquations.jl} package \citep{rackauckas2017dejl}, with absolute and relative tolerances set to $10^{-13}$, to integrate the systems. Each system has its own integration step size, as defined in \texttt{dysts}.
        
        For each system, we collect 7000 data points for training, 3000 data points for validation, and 2500 for testing. The first 300 data points are discarded to account for transient dynamics. The validation points are used for hyperparameter tuning. Each ESN undergoes a grid search to find the optimal value of the ridge regression parameter and leaky coefficient. The hypergrid is determined by latin hypercube sampling \citep{mckay1979comparison, iman1981approach} with boundary values of $[0.1, 1.0]$ for the leaky coefficient, and $[10^{-12}, 10^{-1}]$ for the ridge regression parameter. For training, we provide the MESNs with 10 trajectories per system, each starting from a different location on the attractor. The initial conditions provided in \texttt{dysts} are used to generate the first trajectory. Subsequent trajectories are obtained by shifting the same underlying trajectory by 50 time steps, so that each trajectory begins at a different position on the attractor. 

\section{\label{mmfrc:sec:res} Results}

    In this section we detail the results obtained in this work. To perform the simulations, we used the Julia programming language \citep{bezanson2017julia}. For the integration of the chaotic systems, we relied on \texttt{OrdinaryDiffEq.jl} \citep{rackauckas2017dejl}, while for the visualization we used \texttt{CairoMakie.jl} \citep{danisch2021makie}. All the different reservoir topologies and the implementation, training, and forecasting of the ESNs are sourced through \texttt{ReservoirComputing.jl} \citep{martinuzzi2022reservoircomputing}. For each system pair and topology, we run ten simulations, each starting from a different initial condition. 

    \subsection{\label{mmfrc:subsec:mrsm} Storage vs Selection Across Minimal Deterministic Reservoir Topologies}

        \begin{figure}[ht]
            \centering
            \includegraphics[width=\columnwidth]{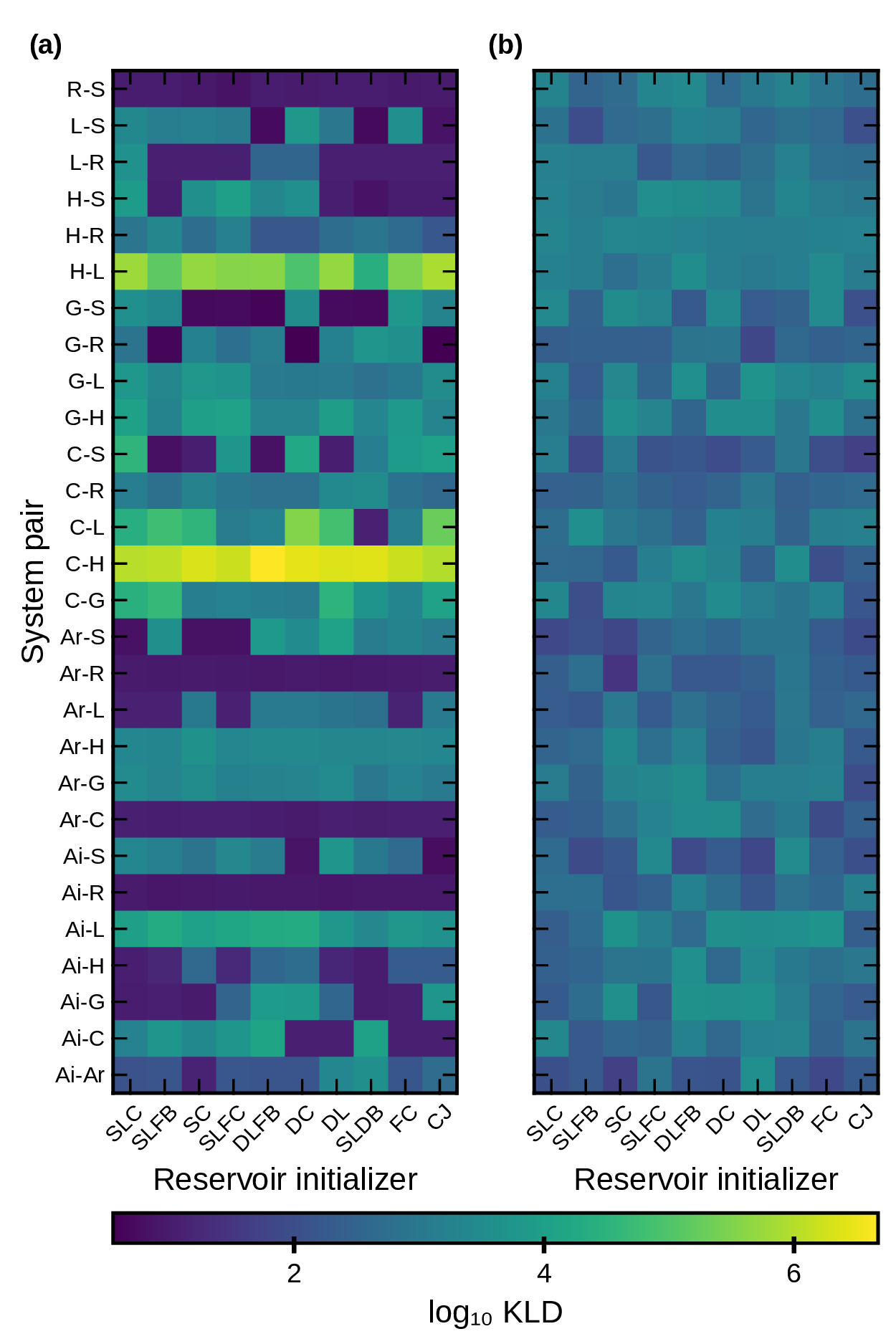}
            \caption{\textbf{Performance of minimal reservoirs across multi-attractor learning tasks and system pairs}. (a) Heat map of prediction quality for each combination of reservoir initializer and chaotic system pair under the blending technique (BT) approach, quantified as the log$_{10}$ of the mean KLD over runs. Panel (b) reports the same results as Panel (a), but for the parameter-aware (PA) training scheme. In both panels, darker tiles indicate lower error (better forecasts), indicating how individual minimal initializers generalize across different blended systems and how their performance patterns change between training schemes.}
            \label{mmfrc:fig:pairheat}
        \end{figure}

        Figure~\ref{mmfrc:fig:mainkld} summarizes the overall performance of the different minimal deterministic reservoir topologies in reproducing multiple chaotic dynamics when trained under the two approaches considered in this study. Performance is reported on a logarithmic scale to accommodate the wide range of observed values. The reported value for the system pair is taken as the average of the two KLDs obtained for each system. The figure aggregates results over all 28 unordered pairs of chaotic systems. In addition to the aggregate error statistics, the figure includes a representative phase-space visualization showing a case in which a single trained model successfully reconstructs both members of a system pair.

        In Fig.~\ref{mmfrc:fig:mainkld}a, we report the distribution of $\log_{10}\mathrm{KLD}$ values obtained in the BT task for each minimal topology. Each light-blue violin corresponds to one topology (CJ, FC, SLDB, DL, DC, DLFB, SLFC, SC, SLFB, SLC; ordered left-to-right by increasing median $\log_{10}\mathrm{KLD}$) and summarizes the variability of errors across all 28 system pairs. The vertical axis extends down to approximately $-2.5$; this lower range is due to the kernel-density smoothing used to draw the violins and does not indicate negative KLD values. The black dot marks the median $\log_{10}\mathrm{KLD}$ for each topology. Under this ordering, the best-performing topology is CJ, with median $\log_{10}\mathrm{KLD}=2.866$ and an interquartile range of $[1.050, 3.543]$ in $\log_{10}\mathrm{KLD}$. On the other side, the worst performing topology is SLC, with median $\log_{10}\mathrm{KLD}=3.357$ and an interquartile range of $[1.095, 3.941]$. Violin width encodes the relative density of outcomes at a given error value. We note that, across all topologies, the distributions are consistently bimodal in log space: using the KDE-based split, the average separation threshold is $\langle \log_{10}\mathrm{KLD}\rangle \approx 1.966$ (std.\ 0.137), and the resulting lobe means averaged over initializers are $\langle \mu_1\rangle = 0.970 \pm 0.064$ and $\langle \mu_2\rangle = 3.642 \pm 0.133$ in $\log_{10}\mathrm{KLD}$. More specifically, for CJ the a KDE-based split at $\log_{10}\mathrm{KLD}\approx 1.973$ separates a lower-error lobe with mean $\log_{10}\mathrm{KLD}=0.929$ from a higher-error lobe with mean $\log_{10}\mathrm{KLD}=3.617$, while for SLC a KDE-based split at $\log_{10}\mathrm{KLD}\approx 2.054$ separates a lower-error lobe with mean $\log_{10}\mathrm{KLD}=1.003$ from a higher-error lobe with mean $\log_{10}\mathrm{KLD}=3.799$.

        In Fig.~\ref{mmfrc:fig:mainkld}b, we report the distribution of $\log_{10}\mathrm{KLD}$ values obtained in the parameter-aware (PA) task for each minimal topology. We keep the topology ordering and visual encoding as in Fig.~\ref{mmfrc:fig:mainkld}. Each violin plot aggregates outcomes across all 28 system pairs, with the black dot indicating the median $\log_{10}\mathrm{KLD}$ for the corresponding topology. Relative to the BT setting, the PA distributions are concentrated in a narrower range, with medians spanning from $2.466$ (CJ) to $3.098$ (DLFB), and interquartile ranges that largely overlap across topologies. Under this ordering, the lowest median is again obtained by CJ (median $\log_{10}\mathrm{KLD}=2.466$, IQR $[2.241, 2.885]$), while the highest median is observed for DLFB (median $\log_{10}\mathrm{KLD}=3.098$, IQR $[2.500, 3.494]$). Unlike the BT results, the violin shapes do not exhibit a consistent shared structure across initializers; instead, the densities vary between topologies and system pairs, reflecting heterogeneous outcomes in the PA setting.

        Finally, we provide a representative phase-space example from the BT experiments in Fig.~\ref{mmfrc:fig:mainkld}c. The panel shows a successful two-attractor reconstruction for the Lorenz and SprottS systems using the SLDB topology. The two systems are shown with distinct colors, and for each system, the reference trajectory is plotted with a dashed line while the corresponding ESN-generated trajectory is plotted with a solid line. The example shown corresponds to a case with $\log_{10}\mathrm{KLD}=0.692$ (i.e., $\mathrm{KLD}\approx 4.92$) for the Lorenz–SprottS pair under SLDB.

        Figure~\ref{mmfrc:fig:pairheat} provides a more detailed view of reconstruction accuracy across the different minimal reservoir topologies by showing the results for the individual system pairs. For each pair of chaotic systems and each reservoir initializer/topology, we report the mean KLD from that experiment, averaged over 10 initial conditions for both systems. Darker colors indicate smaller mean divergences, and thus higher accuracy, while brighter colors indicate larger mean divergences. System pairs are labeled using the first letter of each system name, following the nomenclature defined in Appendix~\ref{mmfrc:app:cs}. For instance, the Lorenz–SprottS pairing is identified as L-S.

        We show the mean KLD heat map for the BT setting in Fig.~\ref{mmfrc:fig:pairheat}a. In the panel, each cell represents the mean of 10 runs from a single BT experiment, defined by a specific system pair and a specific reservoir initializer. We observe pronounced horizontal patterns: for many system pairs, the mean error remains consistently low or consistently high across virtually all investigated topologies. More specifically, we see that the lowest-error rows with consistent performances are Ai–R, with $\langle \log_{10}\mathrm{KLD} \rangle = 0.934$, Ar–R with $\langle \log_{10}\mathrm{KLD} \rangle = 0.970$, and R–S with $\langle \log_{10}\mathrm{KLD} \rangle = 0.979$; Ar–C also displays uniformly low $\log_{10}\mathrm{KLD}$ values across columns (row mean $1.050$). Conversely, the C–H row consistently yields the highest error, with $\langle \log_{10}\mathrm{KLD} \rangle = 6.261$. Following, H–L is the next-highest row with $\langle \log_{10}\mathrm{KLD} \rangle = 5.424$, again showing clear consistency across reservoir designs. In contrast, we find little to no column-wise structure. In fact, no single initializer performs consistently for the majority of system pairs. The CJ topology reports the highest number of successfully reproduced system pairs, consistent with its better performance reported in Fig.~\ref{mmfrc:fig:mainkld}a. Overall, no discernible pattern is evident in the specific behavior of the topologies.

        \begin{figure}[ht]
            \centering
            \includegraphics[width=\columnwidth]{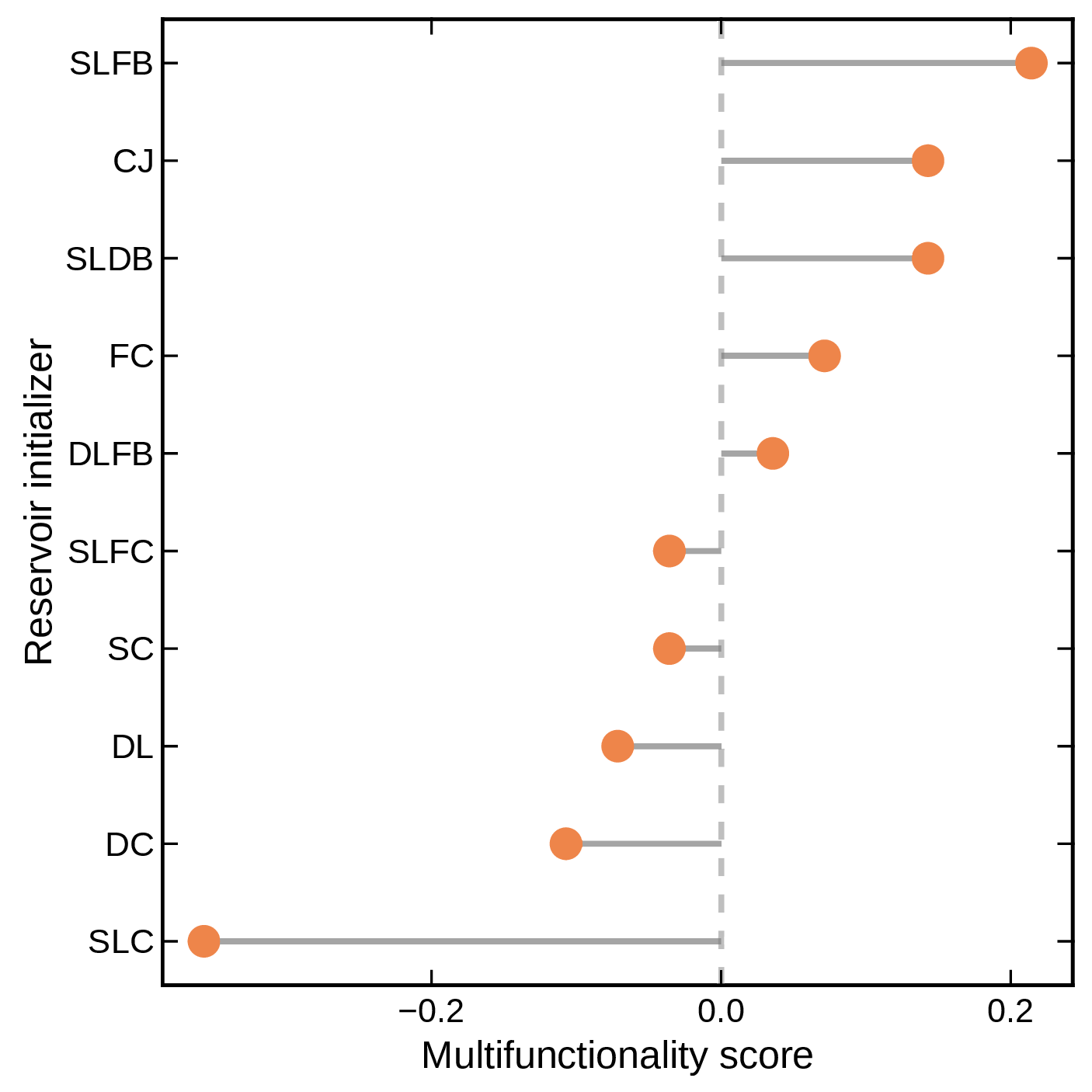}
            \caption{\textbf{Overall multifunctionality ranking of minimal reservoirs}. Multifunctionality score for each reservoir initializer, defined as the difference between the fraction of blended system pairs where the initializer ranks in the top three and the fraction where it ranks in the bottom three (higher is better). Points show the score per initializer, with horizontal lines indicating the distance to zero (vertical dashed line). Initializers at the top with positive scores (e.g., SLFB, CJ, SLDB) are both frequently among the best and rarely among the worst across tasks, while those with negative scores (e.g., SLC) tend to perform poorly more often than well.
            }
            \label{mmfrc:fig:mfscore}
        \end{figure}

        In contrast to panel Fig.~\ref{mmfrc:fig:pairheat}a, we do not find comparably strong structures in Fig.~\ref{mmfrc:fig:pairheat}b, where we report the analogous $\log_{10}$ mean KLD heat map for the PA setting. In this context, the heat map appears more homogeneous, with moderate variation across rows and columns. We observe a single horizontal line corresponding to the H–R pair, which yields the highest errors, with $\langle \log_{10}\mathrm{KLD}\rangle = 3.235$. The wide range of behaviours underlines the results illustrated in Fig.~\ref{mmfrc:fig:mainkld}b, where the PA results also lacked structure compared to the BT counterpart.

    \subsection{\label{mmfrc:subsec:tvs} The role of reservoir topology and system complexity}

        In this Section, we want to further investigate the main drivers of ESN performance in our setting. More specifically, we differentiate between: (i) intrinsic differences in the difficulty of the target system pairs, and (ii) performance differences attributable to reservoir topology. This question arises from the desire to investigate whether certain initializers are broadly effective across many pairings, or if a subset of system pairs is systematically easier to reproduce regardless of the reservoir design. As we can observe from Fig.~\ref{mmfrc:fig:mainkld}a and Fig.~\ref{mmfrc:fig:pairheat}a, the BT setting provides the optimal conditions to investigate this direction. Additionally, the BT approach provides valid reconstructions of the system pairs, whereas PA does not. The PA setting also yields largely overlapping error distributions and weak pair–topology structure in our minimal regime (Figs.~\ref{mmfrc:fig:mainkld}b and \ref{mmfrc:fig:pairheat}b). It would then be difficult to separate topology- versus system-driven effects. Therefore, we use BT as the primary basis for the topology-versus-system analysis in the present section.

        Figure~\ref{mmfrc:fig:mfscore} summarizes an overall ranking of the minimal reservoir initializers using a single “multifunctionality score” computed from the BT experiments. For each initializer (listed on the vertical axis), the score on the horizontal axis is defined as the difference between (i) the fraction of system pairs for which that initializer ranks among the top three performers and (ii) the fraction for which it ranks among the bottom three. It follows that larger values indicate more frequent high rankings and fewer low rankings. Under this definition, we see that SLFB obtains the highest score (0.214), appearing in the top three for 10 of the 28 system pairs and in the bottom three for 4 of the 28 pairs. SLDB and CJ follow with identical scores of 0.143; both initializers appear in the top three around 39\% of the time and in the bottom three for 7/28 pairs. At the opposite end, SLC shows the most negative score, $-0.357$, even though it ranks in the top three for 6/28 pairs and in the bottom three for 16 pairs. The remaining initializers cluster closer to zero. For instance, FC has a small positive score of 0.071, whereas DC (-0.107) and DL (-0.071) lie on the negative side of the zero reference.

\begin{figure}[ht]
            \centering
            \includegraphics[width=\columnwidth]{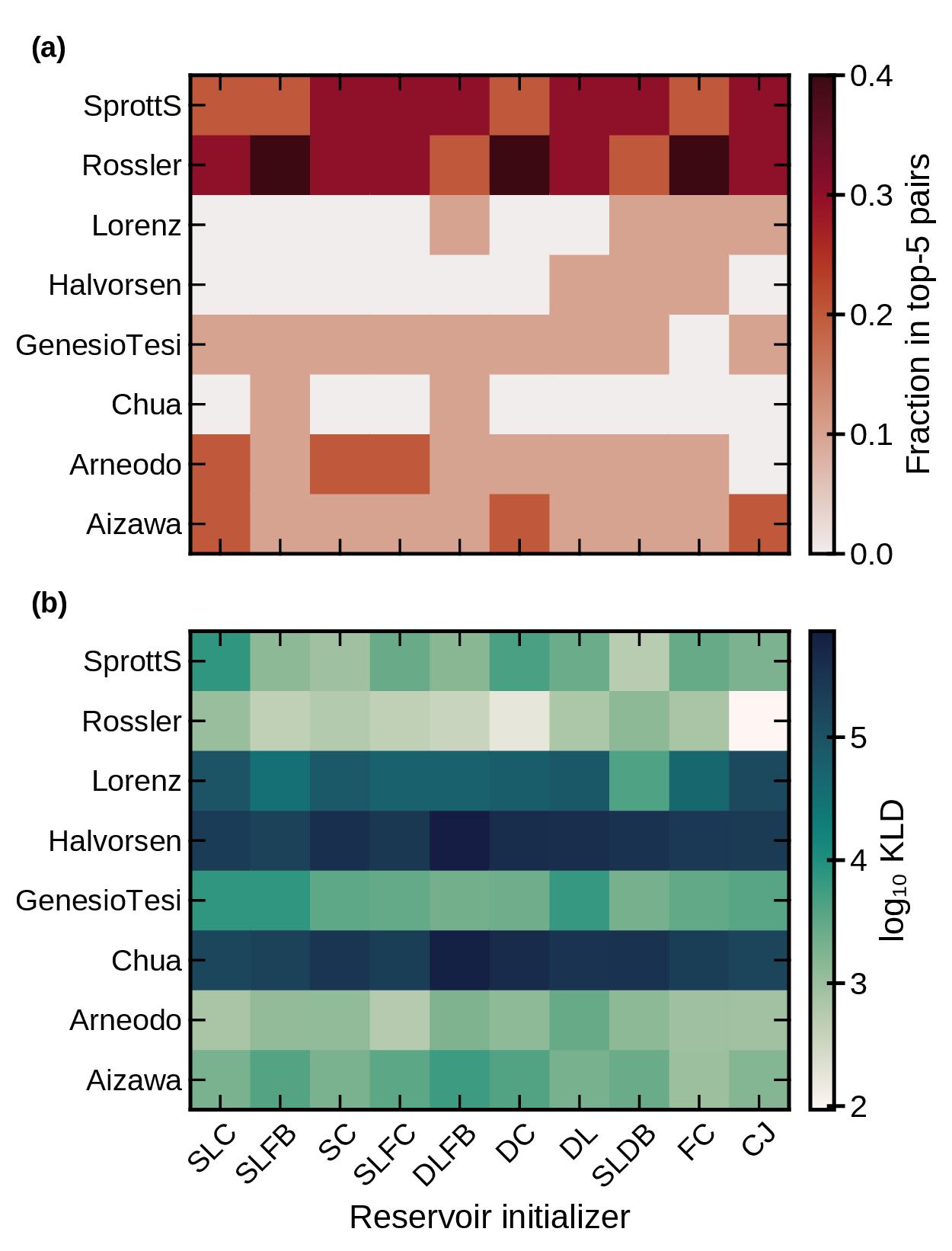}
            \caption{\textbf{System–specific performance of minimal reservoirs}. (a) Fraction of times each chaotic system appears in the top-5 best performing blended pairs for a given initializer (columns). Darker colors indicate that the corresponding initializer tends to perform especially well on tasks involving that system. (b) Mean prediction quality for each system–initializer combination, measured as the log$_{10}$ of the average Kullback-Leibler divergence (KLD) over all blended pairs that contain that system. Lower values (clearer tiles) correspond to better performance. Rows correspond to individual chaotic systems and columns to the minimal reservoir initializers (abbreviations as in the main text).
            }
            \label{mmfrc:fig:cdedist}
        \end{figure}

        Figure~\ref{mmfrc:fig:cdedist} provides an overview of BT performance at the level of individual chaotic systems. We aggregate results across all pairs of blended systems that include a given system, summarizing how each attractor contributes to performance across the set of pairwise tasks. The figure reports two complementary summaries: a rank-based statistic, which measures how frequently each system appears among the best-performing pairs for a given initializer, and an error-based statistic, which reports for each system–initializer combination the $\log_{10}$ of the mean KLD averaged over all pairs that include that system.

        In Fig.~\ref{mmfrc:fig:cdedist}a, we show the rank-based metric. For each initializer, we rank all 28 system pairs by their mean KLD and select the best K=5 pairs. We then count how many times each system appears within these five pairs and divide by 2K=10, since the top-K list contains K pairs and therefore 2K system occurrences in total. Thus, a tile value of 0.40 means that the corresponding system appears 4 times among the 10 system occurrences present in the top-5 pairs (the maximum possible is 0.50, i.e., 5 out of 10, because a system can appear in at most 5 distinct pairs). We find that R\"ossler is the most frequently represented system across initializers, attaining a value of 0.40 for SLFB, DC, and FC, and 0.30 for several others (e.g., SLC, SC, SLFC, DL, CJ). SprottS is also among the top 5 multiple times, achieving 0.30 for SC, SLFC, DLFB, DL, SLDB, and CJ. In contrast, several systems appear only sporadically in the top-5 sets: amongst the worst performers, Lorenz and Halvorsen reach at most 0.10 (e.g., Lorenz: DLFB/SLDB/FC; Halvorsen: DL/SLDB/FC), and Chua appears at most 0.10 (SLFB or DLFB), with some initializers showing 0.00 for that row.

        In Fig.~\ref{mmfrc:fig:cdedist}b, we report the error-based statistic. For each system–initializer combination, we compute the mean KLD averaged over all blended pairs that contain the given system (seven pairs per system) and display its $\log_{10}$ value. The resulting patterns are consistent with the rank-based metric of Fig.~\ref{mmfrc:fig:cdedist}a: systems that appear frequently among the top-$K$ pairs also tend to exhibit lower aggregated errors. For example, the R\"ossler system also shows consistently low values for multiple initializers, ranging from 1.970 (CJ) to 3.136 (SLDB) in average $\log_{10}$ KLD. Likewise, SprottS is repeatedly among the top-$K$ pairs and exhibits comparatively low aggregate errors, ranging from 2.723 (SLDB) to 3.876 (SLC). In contrast, systems that are only weakly represented in panel (a), such as Chua and Halvorsen, occupy the higher ranges in panel (b) (Chua: 5.202–5.829; Halvorsen: 5.281–5.862), reflecting consistently larger mean divergences when averaged across all pairs containing those systems.

\section{\label{mmfrc:sec:disc} Discussion}

    In this work, we show that minimal deterministic reservoir topologies can store multiple chaotic attractors but fail to reliably select among them in response to an external cue. In this work, we evaluated 10 minimal initializers across 28 pairs of chaotic systems. We used two approaches to test the multi-attractor learning ability of each minimal topology: with BT, we assessed the minimal ESNs' ability to store multiple attractors (i.e., their multifunctionality), and with the PA approach, we investigated whether the minimal ESNs could switch between attractors. Across this testbed, we found that in the BT setting, minimal ESNs can store multiple attractors. We did not identify a single topology that consistently performed optimally across all system pairs. Performance is instead largely system-dependent. In particular, pairs involving R\"ossler and SprottS tend to be reproduced with lower divergence, whereas pairs involving Lorenz or Halvorsen more frequently yield larger errors. By contrast, in the minimal-architecture regime studied here, the parameter-aware (PA) setting did not yield reconstructions that met our accuracy criterion for any system pair.

    Why can minimal ESNs learn multiple attractors under the BT setting, yet not under the PA setting for minimal reservoirs? As mentioned in the introduction, BT can be interpreted as a form of content-addressable memory (CAM) \citep{chisvin1992content}, whereas the PA approach more closely resembles location-addressable memory (LAM) \citep{chaudhuri2016computational}. Concretely, the CAM/LAM distinction points to a fundamental difference of the two approaches at the task level: BT primarily tests whether a fixed reservoir can \emph{represent} multiple attractors simultaneously. PA, in contrast, tests whether the reservoir can \emph{select} among attractors based on an external context variable, i.e., whether the external cue can steer the internal state toward the correct basin. In our minimal setting, the systematic failure of PA therefore points to a bottleneck in cue-dependent routing rather than in pure representational capacity. In fact, minimal deterministic reservoirs can sustain multi-attractor structure under BT, but appear to lack sufficient computational power to maintain a stable context signal over the time horizon required for switching under the bias-based labeling of Eq.~\ref{mmfrc:eq:paesn}. Additionally, performance is strongly system-dependent: certain attractors are consistently easier to reproduce than others, as evident in Fig. \ref{mmfrc:fig:pairheat}. A plausible reason for this system dependence is that the benchmark systems differ in dynamical complexity. Using the estimated maximum Lyapunov exponent and entropy-like measures as rough proxies, the systems that are easiest in our BT experiments tend to be less unstable: for example, R\"ossler has a relatively small estimated maximum Lyapunov exponent ($\lambda_{\max}\approx 0.151$) and low Pesin entropy \citep{pesin1977characteristic} ($\approx 0.084$), while SprottS remains moderate ($\lambda_{\max}\approx 0.274$, Pesin $\approx 0.219$). In contrast, the systems that more frequently degrade pairwise performance are characterized by stronger instability and higher entropy rates, e.g., Lorenz ($\lambda_{\max}\approx 0.892$, Pesin $\approx 1.12$) and Halvorsen ($\lambda_{\max}\approx 0.696$, Pesin $\approx 0.586$), with Chua even larger in instability ($\lambda_{\max}\approx 1.23$). These differences suggest that, even under the same training protocol, certain attractors impose a substantially higher effective memory and approximation burden on a fixed minimal reservoir. 
    
    From a design perspective, these observations imply that improving and expanding the learning capability in minimal RC is unlikely to be achieved through small refinements to a single deterministic topology. Instead, PA-style control may require multiple substructures to coexist in the same reservoir. Our interpretation is compatible with the broader idea that multifunctional substrates may decompose into smaller substructures with more specialized roles. While the PA approach does not strictly correspond to multifunctionality as defined in the neuroscience literature, it can nevertheless provide inspiration. For example, \citet{viola2020beyond} discuss how "\textit{a putative multifunctional structure may sometimes be subdivided into smaller structures, each with different functions. So, for instance, the seemingly multifunctional insula turns out to be decomposable into four distinct sub-regions, each one with a more specific function.}" Translating these concepts to reservoir design suggests that composing a reservoir from multiple deterministic sub-reservoirs \citep{li2019echo, oliveirajunior2020clustered, tortorella2024onion, li2025structuring, yoshida2025multiscale} could provide an effective route toward PA-style multi-attractor learning. Such a construction would preserve the determinism and minimal design philosophy explored here, while introducing modular degrees of freedom that may support context retention and cue-dependent routing in more demanding tasks. Relatedly, \citet{dale2020reservoir} argue that while simple topologies can emulate aspects of more complex reservoirs, they often exhibit a restricted behavioral repertoire. In our case, this is consistent with the observation that performance differences across minimal designs do not reduce to a trivial notion of ``more connections is better'': the overall rankings in Fig.~\ref{mmfrc:fig:mfscore} do not align with a simple sparsity, or complexity, ordering, and the limitations of minimal topologies are not addressed by incremental edge additions alone. Another approach would be to tailor reservoirs to a specific group of dynamics \citep{hemmer2024optimal}.

    Our conclusions are conditioned on a deliberately constrained experimental regime. In particular, we train RC models on system pairs whose constituents can exhibit different characteristic time scales. Similar to previous studies \citep{flynn2021multifunctionality, du2025multifunctional}, we use a shared integration step between systems, which helps standardize data generation but does not eliminate cross-system time-scale mismatch. Handling disparate time scales is a known open challenge in RC \citep{tanaka2022reservoir}, and this difficulty is likely amplified in the multi-system setting. We also restrict attention to system pairs drawn from a fixed benchmark of eight fully observed 3D chaotic flows (28 unordered pairs). While this testbed is broader than related studies, the observed patterns may not generalize directly to higher-dimensional dynamics, partial observability, measurement noise, or multi-attractor regimes beyond two systems. Finally, we left the reservoirs unmodified to maintain focus on the topology itself. This choice is consistent with our goal of probing minimality, and in line with prior work \citep{martinuzzi2025minimal}. However, the extent to which optimizing internal weights could alter the attainable performance in minimal RC for multiple attractors remains an open question.

\section{\label{mmfrc:sec:end} Conclusions and Future Directions}
    Minimal complexity reservoirs are desirable for their improved interpretability \citep{hemmer2024optimal} and for potential hardware implementations of RC \citep{appeltant2011information, nakajima2021scalable, abe2024highly}. Using digital substrates, recent studies have demonstrated the strong performance of minimal topologies for learning chaotic dynamics \citep{ma2023novel, ma2023efficient, viehweg2025deterministic, martinuzzi2025minimal}. As the field shifts from single-system forecasting toward models that can operate across regimes, it is timely to ask whether minimal topologies can reproduce multiple dynamical behaviours within a single reservoir. In this study, we show that minimally complex topologies can sustain multistable attractors but cannot switch attractors during the forecasting phase.
    
    Several extensions could clarify which additional ingredients are required for cue-driven switching beyond the minimal regime studied here. First, it would be useful to systematically vary the mechanism by which context is injected into the PA setting (e.g., via a cue, an additional input channel, a learned linear projection, or state augmentation). Second, one can probe whether PA performance is primarily limited by memory capacity by introducing controlled multi-timescale elements \citep{tanaka2022reservoir} (e.g., varying leak rates across units) while keeping connectivity deterministic and interpretable. Third, motivated by the analogy to functional subdivision, modular deterministic reservoirs composed of coupled minimal sub-reservoirs \citep{li2025structuring} provide a natural design axis for testing whether context encoding and attractor representation can be separated structurally.

\begin{acknowledgments}

    The authors acknowledge helpful conversations with Edmilson Roque dos Santos regarding the choice of the error metric and with Francesco Sorrentino regarding the general interpretation of the results. FM also thanks Andrew Flynn for the discussions that helped sharpen our use of the term "multifunctionality".

\end{acknowledgments}

\section*{Author Declaration}

    \subsection*{Conflict of Interest}

        The authors have no conflicts to disclose.

    \subsection*{Author Contributions}

        \textbf{Francesco Martinuzzi}: conceptualization (equal), data curation, formal analysis, methodology, software, validation, visualization, and writing (equal). \textbf{Holger Kantz}: conceptualization (equal), writing (equal).

\section*{Data Availability}

    The data that support the findings of this study are openly available in \url{https://github.com/MartinuzziFrancesco/mmfrc.jl} \citep{martinuzzi2025mmfrc}.

\section*{References}

    \bibliography{mmfrc}

\begin{figure*}[ht]
        \centering
        \includegraphics[width=\textwidth]{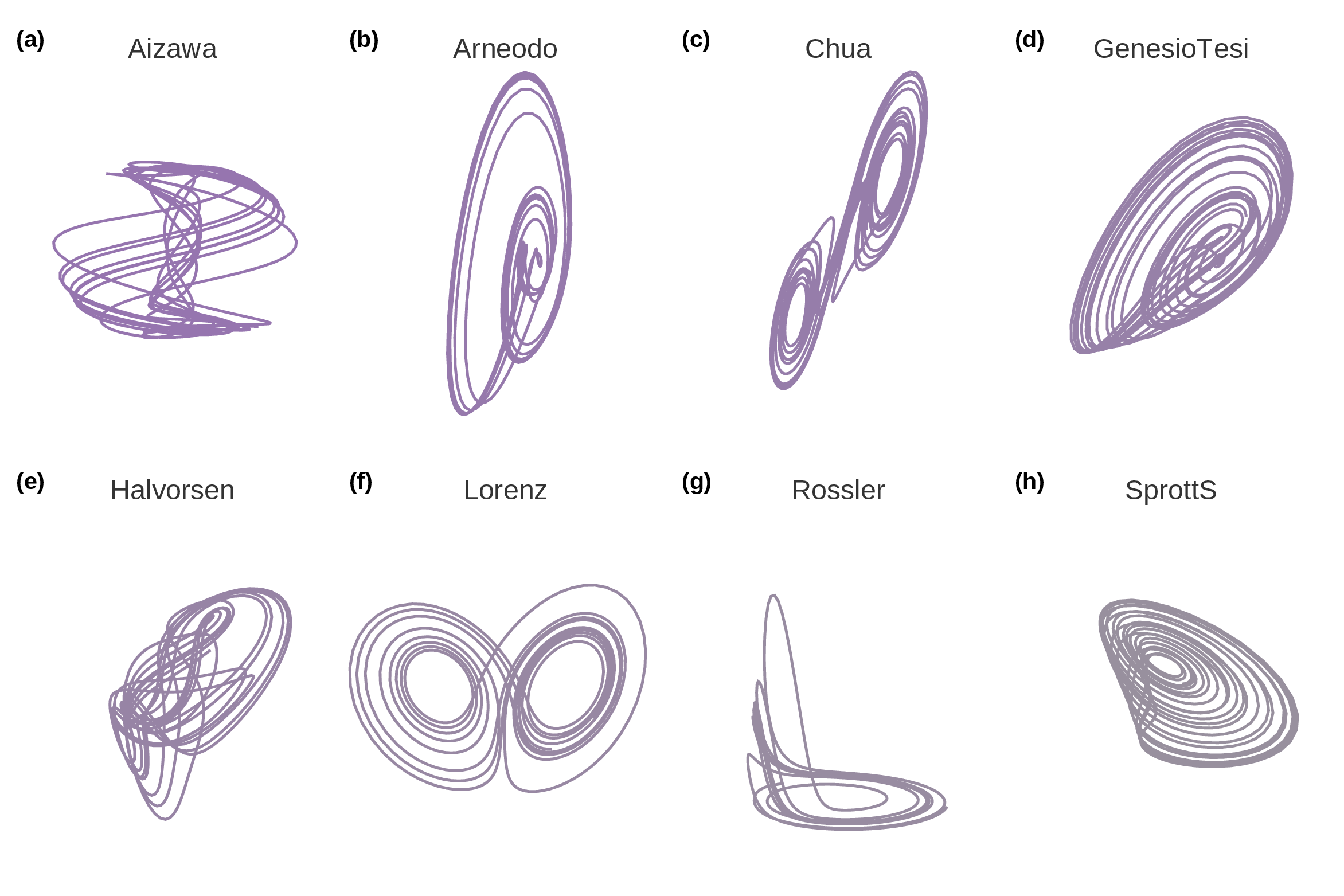}
        \caption{\textbf{Chaotic systems.} Attractors for the eight chaotic systems used in this work: (a) Aizawa, (b) Chen, (c) Chua, (d) GenesioTesi, (e) Halvorsen, (f) Lorenz, (g) R\"ossler, and (h) SprottS.}
        \label{mmfrc:fig:cs}
    \end{figure*}

\appendix

\section{\label{mmfrc:app:cs} Chaotic Systems}

    Here we report the equations, parameters, and integration steps for the eight chaotic systems used in this work. We keep the notation as in \texttt{dysts}. Figure~\ref{mmfrc:fig:cs} shows the corresponding attractors.

    Aizawa \citep{aizawa1982topological} (Fig.~\ref{mmfrc:fig:cs}a)
    \begin{align}
    \label{mmfrc:eq:aizawa}
        \begin{cases}
            \dot{x} &= x z - b x - d y, \\
            \dot{y} &= d x + y z - b y, \\
            \dot{z} &= c + a z - \frac{1}{3} z^{3} - x^{2} - y^{2} - e z x^{2} - e z y^{2} + f z x^{3},
        \end{cases}
    \end{align}
    with parameters $a = 0.95, b = 0.7, c = 0.6, d = 3.5, e = 0.25, f = 0.1$, and integration step $\mathrm{d}t = 9.043 \times 10^{-4}$. 
    
    Arneodo \citep{arneodo1980occurence} (Fig.~\ref{mmfrc:fig:cs}b)
    \begin{equation}
    \label{mmfrc:eq:arneodo}
        \begin{cases}
            \dot{x} = y, \\[0.3em]
            \dot{y} = z, \\[0.3em]
            \dot{z} = -a x - b y - c z + d x^{3},
        \end{cases}
    \end{equation}
    with parameters $a = -5.5, b = 4.5, c = 1.0, d = -1.0$, and integration step $\mathrm{d}t = 6.329 \times 10^{-4}$.
    
    Chua \citep{chua1978nonlinear} (Fig.~\ref{mmfrc:fig:cs}c)
    \begin{equation}
    \label{mmfrc:eq:chua}
        \begin{aligned}
        \begin{cases}
            \dot{x} = \alpha \bigl(y - x - r(x)\bigr), \\[0.3em]
            \dot{y} = x - y + z, \\[0.3em]
            \dot{z} = -\beta y,
            \end{cases}\\[0.6em]
            r(x) &= m_1 x + \dfrac{1}{2}(m_0 - m_1)\bigl(\lvert x + 1\rvert - \lvert x - 1\rvert\bigr),
        \end{aligned}
    \end{equation}
    with parameters $\alpha = 15.6, \beta = 28.0, m_{0} = -1.142857, m_{1} = -0.71429$, and integration step $\mathrm{d}t = 6.605 \times 10^{-3}$.
    
    GenesioTesi \citep{genesio1992harmonic} (Fig.~\ref{mmfrc:fig:cs}d)
    \begin{equation}
    \label{mmfrc:eq:genesiotesi}
        \begin{cases}
            \dot{x} = y, \\[0.3em]
            \dot{y} = z, \\[0.3em]
            \dot{z} = -c x - b y - a z + x^{2},
        \end{cases}
    \end{equation}
    with parameters $a = 0.44, b = 1.1, c = 1$, and integration step $\mathrm{d}t = 2.508 \times 10^{-3}$.

    Halvorsen \citep{sprott2010elegant} (Fig.~\ref{mmfrc:fig:cs}e)
    \begin{equation}
    \label{mmfrc:eq:halvorsen}
        \begin{cases}
            \dot{x} = -a x - b y - b z - y^{2}, \\[0.3em]
            \dot{y} = -a y - b z - b x - z^{2}, \\[0.3em]
            \dot{z} = -a z - b x - b y - x^{2},
        \end{cases}
    \end{equation}
    with parameters $a = 1.4, b = 4$, and integration step $\mathrm{d}t = 2.144 \times 10^{-4}$.

    Lorenz \citep{lorenz1963deterministic} (Fig.~\ref{mmfrc:fig:cs}f)
    \begin{equation}
    \label{mmfrc:eq:lorenz}
        \begin{cases}
            \dot{x} = \sigma (y - x), \\[0.3em]
            \dot{y} = x (\rho - z) - y, \\[0.3em]
            \dot{z} = x y - \beta z,
        \end{cases}
    \end{equation}
    with parameters $\beta = 2.667, \rho = 28, \sigma = 10$, and integration step $\mathrm{d}t = 1.801 \times 10^{-4}$.

    R\"{o}ssler \citep{rossler1976equation} (Fig.~\ref{mmfrc:fig:cs}g)
    \begin{equation}
    \label{mmfrc:eq:rossler}
        \begin{cases}
            \dot{x} = -y - z, \\[0.3em]
            \dot{y} = x + a y, \\[0.3em]
            \dot{z} = b + x z - c z,
        \end{cases}
    \end{equation}
    with parameters $a = 0.2, b = 0.2, c = 5.7$, and integration step $\mathrm{d}t = 7.563 \times 10^{-4}$.

    SprottS \citep{sprott1994some} (Fig.~\ref{mmfrc:fig:cs}h)
    \begin{equation}
    \label{mmfrc:eq:sprotts}
        \begin{cases}
            \dot{x} = -x - 4y, \\[0.3em]
            \dot{y} = x + z^{2}, \\[0.3em]
            \dot{z} = 1 + x,
        \end{cases}
    \end{equation}
    with integration step $\mathrm{d}t = 1.3 \times 10^{-3}$.

\end{document}